\documentclass[12pt,times]{article}
\title{Bridging scales in multiscale bubble growth dynamics with correlated fluctuations using neural operator learning}

\author{Minglei Lu$^{1*}$, Chensen Lin$^2$\footnote{The first two authors contributed equally to this work}, Martian Maxey$^3$,\\ George Em Karniadakis$^3$ and Zhen Li$^1$\footnote{Email: \href{mailto:zli7@clemson.edu}{zli7@clemson.edu}}\\
\small{$^1$ Department of Mechanical Engineering, Clemson University, Clemson, SC 29634, USA}\\
\small{$^2$ Artificial Intelligence Innovation and Incubation Institute, Fudan University, Shanghai, 200433, China}\\
\small{$^3$ Division of Applied Mathematics, Brown University, Providence, RI 02912, USA}}

\date{}

\RequirePackage{lineno}
\usepackage[colorlinks=true]{hyperref}
\usepackage{graphicx}
\usepackage{multirow,array,color,mathrsfs,subcaption}
\usepackage{epstopdf}
\usepackage{caption}

\DeclareGraphicsExtensions{.eps,.mps,.pdf,.jpg,.png}
\usepackage[sort&compress,numbers]{natbib}
\usepackage{fancyhdr}
\usepackage{amsmath}
\usepackage{pxfonts}

\usepackage[top=1.5cm,bottom=2.5cm,left=1.5cm,right=1.5cm]{geometry}

\usepackage{amsmath,amssymb}
\begin{document}

\maketitle
\vspace{-15pt}
\begin{abstract}
The intricate process of bubble growth dynamics involves a broad spectrum of physical phenomena from microscale mechanics of bubble formation to macroscale interplay between bubbles and surrounding thermo-hydrodynamics.
Traditional bubble dynamics models including atomistic approaches and continuum-based methods segment the bubble dynamics into distinct scale-specific models. In order to bridge the gap between microscale stochastic fluid models and continuum-based fluid models for bubble dynamics, we develop a composite neural operator model to unify the analysis of nonlinear bubble dynamics across microscale and macroscale regimes by integrating a many-body dissipative particle dynamics (mDPD) model with a continuum-based Rayleigh-Plesset (RP) model through a novel neural network architecture, which consists of a  deep operator network for learning the mean behavior of bubble growth subject to pressure variations and a long short-term memory network for learning the statistical features of correlated fluctuations in microscale bubble dynamics. Training and testing data are generated by conducting mDPD and RP simulations for nonlinear bubble dynamics with initial bubble radii ranging from 0.1 to 1.5 micrometers. Results show that the trained composite neural operator model can accurately predict bubble dynamics across scales, with a 99\% accuracy for the time evaluation of the bubble radius under varying external pressure while containing correct size-dependent stochastic fluctuations in microscale bubble growth dynamics. The composite neural operator is the first deep learning surrogate for multiscale bubble growth dynamics that can capture correct stochastic fluctuations in microscopic fluid phenomena, which sets a new direction for future research in multiscale fluid dynamics modeling.
\end{abstract}

\section{Introduction}\label{sec:1}
Bubbles play a crucial role in numerous industrial processes, such as water treatment~\cite{2023_Jia_Nanobubbles_WR}, mineral flotation~\cite{2020_Wang_Regulation_JECE}, microrobotics~\cite{2022_Zhou_Review_Micromachines}, propeller cavitation~\cite{2023_Wang_Anticorrosive_POC}, and the food and beverage industry~\cite{2023_Zhang_Micro_FRI}.
The nonlinear dynamics of bubble growth is inherently a multiscale phenomenon that involves a wide range of physical processes and scales in the bubble formation and growth, as well as the intricate interplay between bubbles and surrounding thermo-hydrodynamics. At the continuum limit, the bubble growth dynamics can be described by the partial differential equations for conservation of mass and conservation of momentum. Under the assumption of spherical symmetry, substantial knowledge has been built up over the past several decades about bubble dynamics through Rayleigh-Plesset (RP) models~\cite{2017_Prosperetti_Vapor_ARFM}. Considering a spherical bubble in an infinite body of incompressible Newtonian fluid, the evolution of the bubble radius can be determined by the Rayleigh-Plesset equation~\cite{2017_Prosperetti_Vapor_ARFM},
\begin{equation}
\frac{P_B(t)-P_\infty(t)}{\rho_L}=R\frac{d^2R}{dt^2}+\frac{3}{2}\left(\frac{dR}{dt}\right)^2 + \frac{4\nu_L}{R}\frac{dR}{dt}+\frac{2\sigma_{LV}}{\rho_L R},
\label{eq:RP}
\end{equation}
where the fluid properties, including the liquid density $\rho_L$, the kinematic viscosity of the surrounding liquid $\nu_L$, and the liquid-vapor surface tension $\sigma_{LV}$, are assumed to be constant. Provided that the bubble pressure $P_B(t)$ and the external pressure $P_\infty(t)$ in the far field are given, the time evolution of bubble radius $R(t)$ can be directly solved by Eq.~(\ref{eq:RP}), which has been used in understanding bubble dynamics in various engineering applications. Examples includes the investigation of the bubble dynamics in ultrasound fields to improve the efficacy of ultrasound-based diagnostics and therapeutics~\cite{2013_Doinikov_Ultra_PMB},
the study of acoustic droplet vaporization and its application in gas embolotherapy by employing the RP equation for bubble dynamics~\cite{2013_Qamar_Dynamics_APL}, and incorporation of the RP equation into large eddy simulations to consider the impact of cavitation and bubbles in modifying turbulent flow characteristics~\cite{2015_Ji_Large_IJMF}.

\begin{figure}[bh!]
    \centering
    \includegraphics[width=0.35\textwidth]{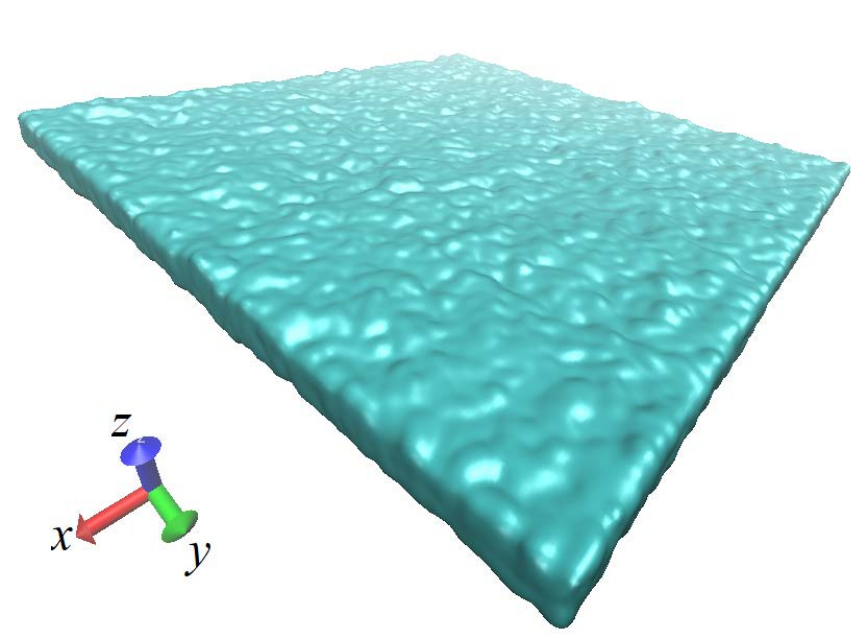}
    \hspace{1cm}
    \includegraphics[width=0.4\textwidth]{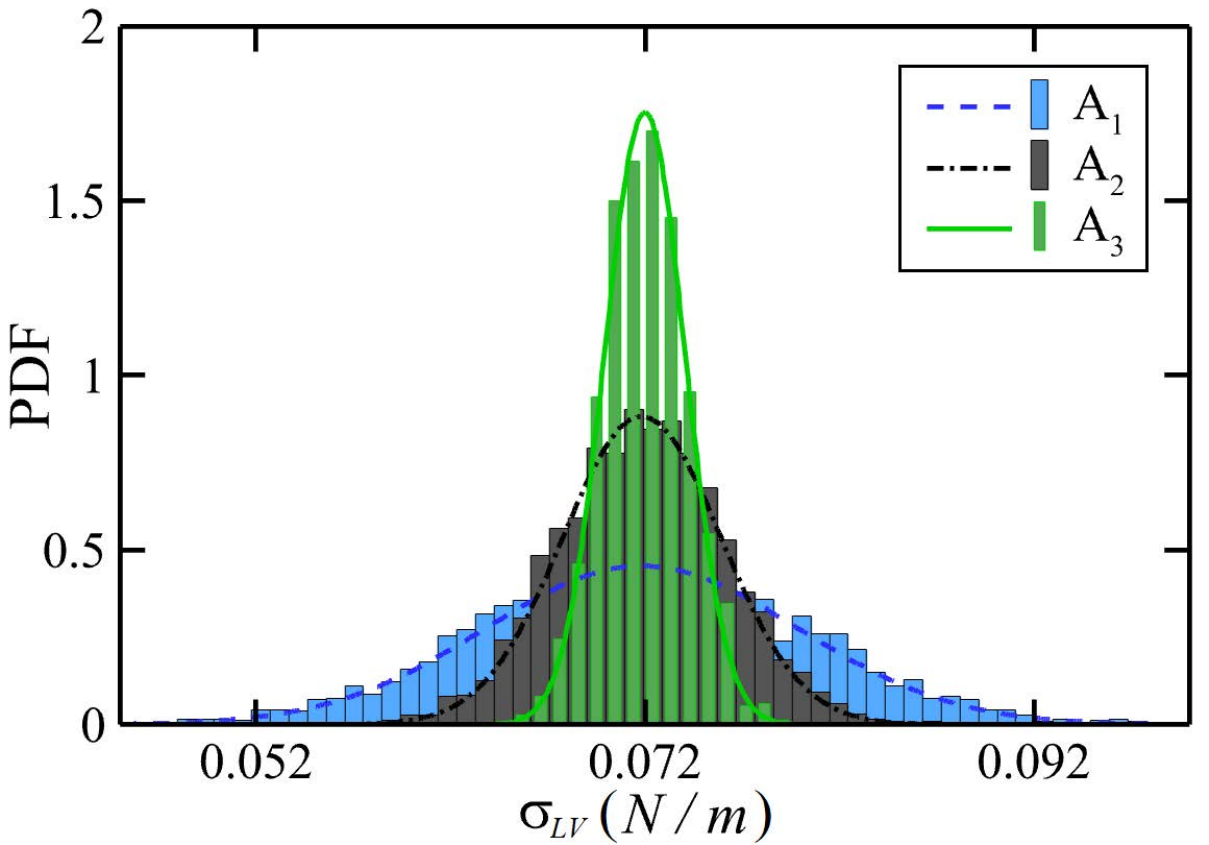}\\
    (a)\hspace{7cm}(b)
    \caption{Interfacial tension at microscale. (a) Setup of a multiphase dissipative particle dynamics simulation of a thin liquid film for microscale surface tension measurement, and (b) the probability density function (PDF) of instantaneous liquid-vapor surface tension computed from the liquid film with different surface areas.}
    \label{fig-film}
\end{figure}

At the microscale, the bubble growth dynamics is determined by the competition between bulk energy and interfacial tension subject to fluctuating hydrodynamics~\cite{2020_Gallo_Nucleation_JFM}, leading to a strongly stochastic process. Moreover, the fluid properties such as the kinematic viscosity and the interfacial tension in microscale fluctuating hydrodynamics in equilibrium are in general normally distributed random variables rather than constant quantities used in the continuum-based RP models.
Figure~\ref{fig-film}(a) presents a setup of a multiphase dissipative particle dynamics (DPD) simulation of a thin liquid film at rest, where the liquid-vapor interfacial tension is measured by the difference between the normal and tangential stress integrated across the interface~\cite{2023_Sheikh_Brownian_JCP},
\begin{equation}
\sigma_{LV}=\frac{1}{2A}\int_V\left(p_{zz}-\frac{p_{xx}+p_{yy}}{2}\right)dV,
\label{eq:surface tension}
\end{equation}
where $\sigma_{LV}$ represents the liquid-vapor interfacial tension, $A$ is the surface area of the upper side of the liquid film, $p_{xx}$, $p_{yy}$, and $p_{zz}$ are the stress components along $x$, $y$ and $z$-directions, respectively. The factor $1/2$ in Eq.~(\ref{eq:surface tension}) takes account for the fact that the film has two liquid-vapor interfaces (the upper and lower surfaces). Figure~\ref{fig-film}(b) shows the probability density function (PDF) of instantaneous liquid-vapor surface tension computed from three squared liquid films with different surface areas, i.e., $A_1=0.04~{\rm \mu m}^2$, $A_2=0.16~{\rm \mu m}^2$, and $A_3=0.64~{\rm \mu m}^2$. The computed surface tensions have the same mean value for all three surface areas, which is $0.072~{\rm N/m}$. However, the surface tension computed from the smallest surface area $A_1$ spreads out over a wider range of values, which indicates that the variance of the instantaneous surface tension is inversely proportional to the surface area of the liquid film, with an expectation that the variance approaches to zero when the surface area is very large at the continuum limit, but it could be comparable to the surface tension at small length scales.
Therefore, at the microscale, i.e., nanoscale and submicron bubbles, stochastic fluid models such as direct molecular dynamics (MD)~\cite{2006_Nagayama_Molecular_IJHMT} and many-body dissipative particle dynamics (mDPD)~\cite{2018_Pan_Mesoscopic_EL} are employed to study the role of hydrodynamic fluctuations in determining bubble shape and growth dynamics.

In order to predict the full-scale bubble growth dynamics from microscale to continuum scale, the most common practice is to simulate the microscale bubble dynamics using stochastic fluid models such as MD and mDPD and the macroscale bubble dynamics using continuum-based fluid models~\cite{2022_Sullivan_Inertio_JFM}.
Large-scale MD simulations can be performed to investigate the collapse of a bubble in liquid water, but the system size is limited to a length up to $0.5~{\rm \mu m}$ containing $4.5\times10^9$ molecules because of the high computational cost~\cite{2023_Chen_Large_JCP}. Continuum-based computational fluid dynamics simulations including the volume of fluid method~\cite{2022_Mulbah_Areview_PNE} and the level-set method~\cite{2015_Balcazar_Level_IJHF} can be used to study macroscale bubble growth dynamics. Also, a hybrid atomistic-continuum method~\cite{2017_Zhang_Hybrid_NHT} has been developed to conduct a multiscale investigation on bubble dynamics by integrating MD simulations for the detailed examination of bubble growth at the microscale with continuum-based computational fluid dynamics for capturing the fluid flow and heat transfer at the macroscale. However, these approaches have to manually split a single physical process into different length/time regions to be studied using standalone physics-based fluid models. Recent studies by Lin et al.~\cite{2021_Lin_Operator_JCP,2021_Lin_ASeamless_JFM} demonstrated that an operator regression framework based on the deep operator network (DeepONet) can act as a unified surrogate model to seamlessly bridge atomistic and continuum scales in multiscale bubble growth dynamics, where the fluctuations in microscale bubble dynamics were eliminated by ensemble averaging multiple mDPD simulations to predict the mean bubble growth dynamics. However, the stochastic fluctuation in microscale bubble dynamics is not a Gaussian white noise, which is correlated in time through intricate interplay between the bubble dynamics and fluctuating hydrodynamics. It represents the inherent stochasticity of the fluid system at the microscale and should be captured by the predictive bubble dynamics model.

In this paper, we aim to develop a composite neural operator model as surrogate of physics-based bubble dynamics models to unify the continuum and stochastic regimes, with the capability of predicting correct correlated fluctuations in nonlinear bubble dynamics at the microscale. The reminder of this paper is organized as follows. In Section~\ref{sec:2}, we introduce the details of the microscopic model for the bubble dynamics in stochastic regime, and the macroscale R-P model for the bubble dynamics in continuum regime. In Section~\ref{sec:3}, we develop a composite neural operator network by integrating a long short-term memory network into the branch net of DeepONet to learn the statistical features of bubble dynamics. Section~\ref{sec:4} presents the results and the performance analysis of the composite neural operator network in predicting both microscale and macroscale bubble dynamics. Finally, we conclude this paper with a brief summary and discussion in Section~\ref{sec:5}.

\section{Multiscale Modeling of Bubble Growth Dynamics}\label{sec:2}
\subsection{Microscale Stochastic Model}\label{sec:2.1}
We employ a multiphase DPD model to simulate the microscale bubble growth dynamics. DPD is a coarse-grained MD approach that can model stochastic fluid dynamics associated with correct fluctuation correlations~\cite{2017_Espanol_Perspective_JCP}. The governing equation of DPD can be rigorously derived by applying the Mori-Zwanzig projection to atomistic dynamics~\cite{2014_Li_Construction_SM}, which builds a direct connection between DPD and MD. In addition, the mean-field hydrodynamic equations of a DPD system recover the Navier-Stokes equations in the continuum limit~\cite{1997_Marsh_Static_PRE}, which allows a smooth transition between microscale bubble dynamics and continuum bubble dynamics. The equation of motion for a DPD particle $i$ follows the Newton's equation of motion~\cite{groot1997dissipative},
%
\begin{equation}    \label{eq.DPD.EOM}
m_i\frac{{\rm d}^2\mathbf{r}_i}{{\rm d}t^2} = m_i\frac{{\rm d}\mathbf{v}_i}{{\rm d}t} = \mathbf{F}_i = \sum_{j\ne i}\left(\mathbf{F}^C_{ij} + \mathbf{F}^D_{ij} + \mathbf{F}^R_{ij}\right),
\end{equation}
%
where $m_i$ represents the mass of a particle $i$, $\mathbf{r}_i$ and $\mathbf{v}_i$ are position and velocity vectors of the particle $i$, and $\mathbf{F}_i$ is the total force acting on the particle $i$ owing to the interactions with neighboring particles.
The computation of $\mathbf{F}_i$ involves a summation over all neighboring particles within a specified cutoff range.
The pairwise force encompasses a conservative force $\mathbf{F}^C_{ij}$, a dissipative force $\mathbf{F}^D_{ij}$ and a random force $\mathbf{F}^R_{ij}$. These forces are given in the following forms:
%
\begin{equation}    \label{eq.DPD.force}
\begin{split}
   & \mathbf{F}^C_{ij} = a_{ij}\omega_C(r_{ij})\mathbf{e}_{ij}, \\
   & \mathbf{F}^D_{ij} = -\gamma_{ij}\omega_D(r_{ij})(\mathbf{v}_{ij}\mathbf{e}_{ij})\mathbf{e}_{ij}, \\
   & \mathbf{F}^R_{ij}\cdot {\rm d}t = \sigma_{ij}\omega_R(r_{ij}){\rm d}\tilde{W}_{ij}\mathbf{e}_{ij},
\end{split}
\end{equation}
%
where $r_{ij}=|\mathbf{r}_{ij}| = |\mathbf{r}_i-\mathbf{r}_j|$ represents the distance between particles $i$ and $j$, $\mathbf{e}_{ij}=\mathbf{r}_{ij}/r_{ij}$ is the unit vector, $\mathbf{v}_{ij} = \mathbf{v}_i-\mathbf{v}_j$ is the velocity difference.
Additionally, ${\rm d}\tilde{W}_{ij}$ stands for an independent increment of the Wiener process~\cite{espanol1995statistical}.
The coefficients $a_{ij}$, $\gamma_{ij}$ and $\sigma_{ij}$ play crucial roles in determining the strength of the conservative, dissipative, and random forces, respectively.
To adhere to the fluctuation-dissipation theorem ~\cite{espanol1995statistical}, and maintain the DPD system at a constant temperature~\cite{groot1997dissipative}, constraints are imposed on dissipative and random forces. Specifically, the constraints are expressed by $\sigma_{ij}^2 = 2\gamma_{ij}k_BT$ and $\omega_D(r_{ij})=\omega^2_R(r_{ij})$, ensuring a consistent and thermodynamically valid representation of the forces in the DPD model.

\begin{figure}[t]
    \centering
    \includegraphics[width=0.70\textwidth]{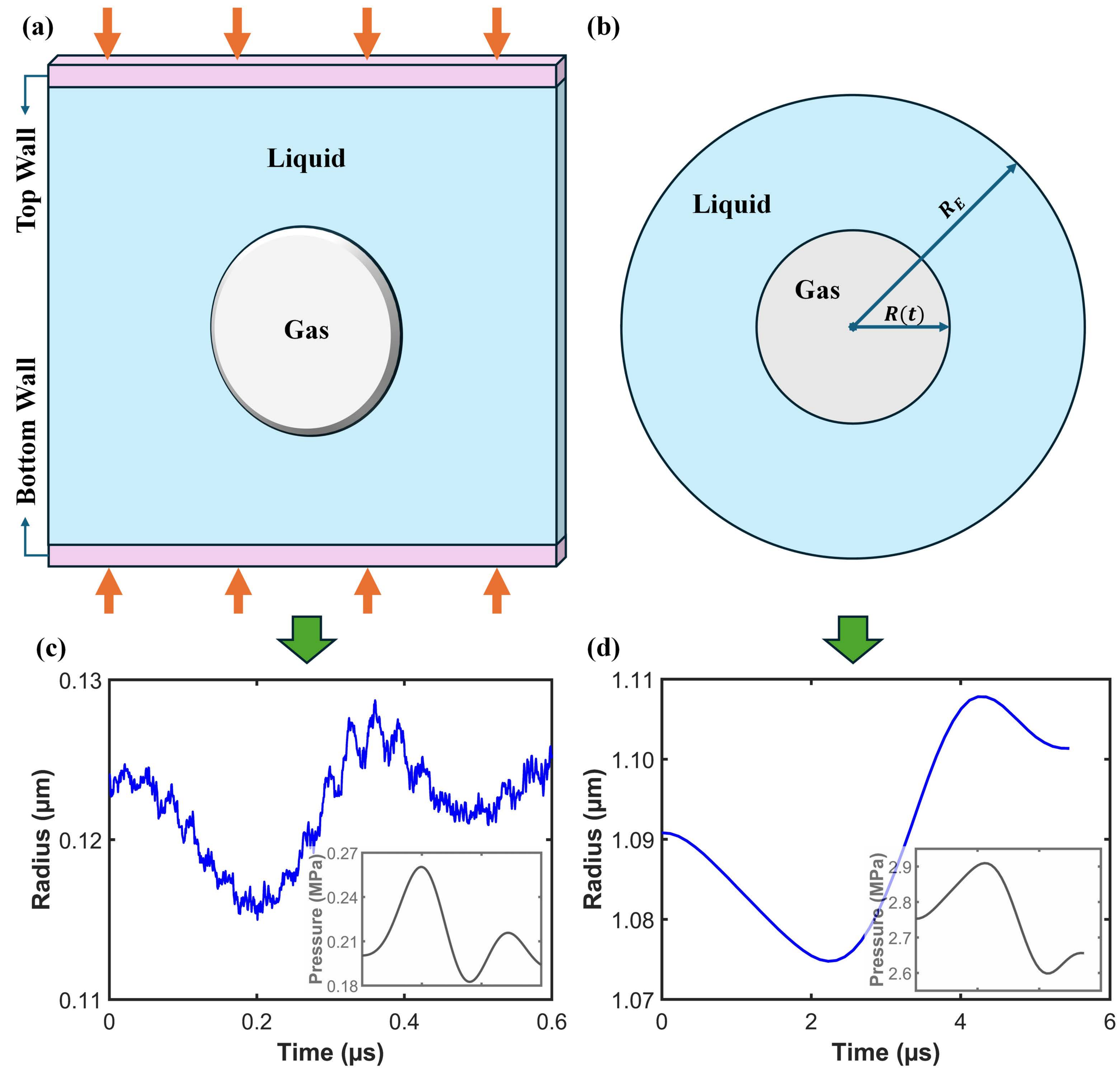}
    \caption{(a) The DPD-MDPD model at the microscale. The grey gas phase uses DPD simulation and the blue liquid phase uses MDPD simulation. The top and bottom walls can move with time-varying pressure. (b) The R-P model for a single bubble in liquid at the macroscale with time-dependent pressure change at the boundary at $R_E$. (c) One case of DPD-MDPD simulation results: Bubble radius $R(t)$ changes with pressure changes from initial bubble radius $R_0=124\rm{nm}$. (d) One case of R-P simulation results: Bubble radius $R(t)$ changes with pressure changes from initial bubble radius $R_0=1.09\rm{\mu m}$. The inset shows the pressure change.}
    \label{fig-simulation-setup}
\end{figure}

DPD, characterized by purely repulsive conservative forces, is well-suited for simulating gaseous systems characterized by spontaneous spatial filling.
To capture the intricacies of liquid-phase behavior, the many-body DPD (mDPD) extension is employed, altering the conservative force to incorporate both attractive and repulsive ~\cite{warren2003vapor,arienti2011many},
\begin{equation}    \label{eq.MDPD.force}
\mathbf{F}_{ij}^C = A_{ij}w_c(r_{ij}) + B_{ij}(\rho_i + \rho_j) w_d(r_{ij}).
\end{equation}
%
To address stability concerns at the interface, where a single interaction range proves inadequate~\cite{pagonabarraga2001dissipative}, the repulsive contribution is confined to a shorter range $r_d < r_c$ compared to the soft pair attractive potential.
The many-body repulsion, expressed as a self-energy per particle, takes a quadratic form in local density, $B_{ij}(\rho_i+\rho_j)w_d(r_{ij})$, where $B_{ij}>0$.
The density of each particle is defined as $\rho_i = \sum_{j \neq i}w_{\rho}(r_{ij})$,
and its weight function $w_{\rho}$ is defined as $w_{\rho}={15}/({2 \pi r_d^3})\cdot\left(1-{r}/{r_d}\right)^{2}$ in which $w_\rho$ vanishes beyond a cutoff distance $r_d$.

In Figure~\ref{fig-simulation-setup}(a), we present the illustration of the DPD-mDPD coupled simulation system. Here, DPD potentials are used for gas phase and mDPD potentials for liquid phase.
The model has been validated by comparing it to the RP equation for the larger size bubbles and with MD for the smaller size bubbles~\cite{pan2018mesoscopic}.
For computational efficiency, a thin box with periodic boundaries in the $x$ and $z$ directions is employed.
To manipulate the system pressure, the top and bottom walls possess one degree of freedom, allowing movement solely in the $y$ direction.
The external force (orange arrows) on the walls follows a predetermined function.
We continuously monitor the volume of the gas phase and use it to compute the Continuous monitoring of the gas phase volume is performed, and Voronoi tessellation~\cite{rycroft2009voro} is employed to estimate the instantaneous volume occupied by gas particles, thereby allowing computation of the effective bubble radius.
To generate enough data of microscale bubble growth dynamics for the neural operator training, we adopt a well tested fluid system for bubble dynamics developed by Lin et al.~\cite{2021_Lin_Operator_JCP} with the DPD and mDPD parameters listed in its Table II.

\subsection{Macroscale Continuum Model}\label{sec:2.2}

The standard Rayleigh-Plesset equation for a spherical bubble is derived from the continuum-level Navier-Stokes equations~\cite{plesset1977bubble}.
It serves to depict the variation in radius $R(t)$ of a single, spherical gas-vapor bubble in liquid as the far-field pressure, $p_{\infty}(t)$, undergoes changes over time $t$.
To align with the mean-field dynamics of the DPD system, we introduce a new two-dimensional (2D) version of the RP model for a gas bubble within a finite liquid domain, featuring a finite gas-liquid density ratio.
The conceptual framework of this model is illustrated in Figure~\ref{fig-simulation-setup}(b), showing a circular gas bubble with radius $R$ inside an external circular boundary of radius $R_{E}$.
The presence of the external boundary is imperative due to the absence of a finite limit for the fluctuating pressure as $r \to \infty$.
Here, $R_{E}$ is determined by ensuring the matching the liquid volume between the RP system and the DPD system.

Initially, we focus on the motion in the surrounding liquid, presuming it to be incompressible, characterized by a constant density $\rho_L$ and viscosity $\mu_L$.
During the bubble expansion, a radial potential flow $u(r, t)$ emerges, designed to adhere to the kinematic conditions at both $r = R$ and $r = R_E$ such that
\begin{eqnarray}
 u(R, t)    &=&   \frac{\mathrm{d} R}{\mathrm{d} t} \\
 u(R_E,t) &=&   \frac{\mathrm{d} R_E}{\mathrm{d} t} = \frac{R}{R_E} \frac{\mathrm{d} R}{\mathrm{d} t}.
\end{eqnarray}
The latter equation determines $R_E (t)$ and ensures the constancy of $\pi (R_E^2 - R^2) = V_L$, where $V_L$ represents the liquid volume per unit length in $z$ direction.
The liquid pressure,derived from the Navier-Stokes equation with $u=R/r \cdot \mathrm{d}R/\mathrm{d}t$ (up to an arbitrary function $g(t)$), is given by:
\begin{equation}
    -p_{L} (r,t) = \rho_{L} \left \{ \frac{\mathrm{d}}{\mathrm{d}t} \left( R\frac{\mathrm{d}R}{\mathrm{d}t} \right) \log r + \frac{u^2}{2} \right \} + g(t).
\end{equation}
The liquid pressure at $r=R_E$ is specified as $p_{L} (R_E,t) = p_{E} (t)$.
At $r=R$, the difference in the normal stresses balances the surface tension,
\begin{equation}
    \left \{ -p_{L} + 2\mu_{L} \frac{\partial u}{\partial r} \right \} \bigg|_{R+} + \frac{\gamma}{R} = \tau^B_{rr}(t),
\end{equation}
where $\gamma$ is the coefficient of surface tension at the gas-liquid interface and $\tau^B_{rr}(t)$ is the normal stress in the gas phase at the bubble surface.

When neglecting the inertia and the viscous stresses of the gas phase, with the only contribution to the normal stress being the gas pressure, $\tau^B_{rr}(t) = -p_B(t)$, the resulting 2D RP model for a circular gas bubble is expressed as:
\begin{eqnarray}
p_B(t) - p_{L} (R_E,t)  =  \rho_{L}\log_{e}\left(\frac{R_E}{R}\right)\frac{d}{dt}\left(R\frac{dR}{dt}\right)
- \frac{1}{2}\rho_L\left(1-\frac{R^2}{R_{E}^2}\right)\left(\frac{dR}{dt}\right)^2
 + 2 \mu_L \frac{1}{R}\frac{dR}{dt} +  \frac{\gamma}{R}.
\label{eq.RP}
\end{eqnarray}

Inside the bubble, the thermodynamic pressure of the gas follows a polytropic gas law:
\begin{equation}
{p_{B}(t) = p_{G0} \left(\frac{T_B}{T_\infty}\right)
\left(\frac{R_0}{R}\right)^{k}},
\label{eq.gas}
\end{equation}
where $k$ is an approximately constant parameter associated with the system state.
Through a series of quasi-static DPD simulations conducted at various liquid pressure, the value of $k$ was calibrated.
The best-fit value obtained from the simulation data over the working range was $k = 3.68$. In these simulations, a thermostat was employed to ensure stability, rendering the system approximately isothermal.
While the temperature of an oscillating bubble can be influenced by factors like viscous heating or pressure work, this study assumes that the fluid system is connected to a thermostat bath, maintaining a constant system temperature.
Consequently, we neglect the impact of temperature differences between the gas and the liquid, setting $T_{B} = T_{\infty}$.
In both the DPD simulations and the RP model, the initial conditions prescribe that the gas bubble is in thermal equilibrium at $t = 0$ with an initial radius of $R_0$, a gas pressure $p_{G0}$, and a liquid pressure $p_{L}(0)=p_{E}(0)$.
The initial gas and liquid pressures adhere to the 2D Young-Laplace equation:
\begin{equation}  \label{eq.initial.condition}
	{p_{G0} = p_{L}(0)+\frac{\gamma}{R_{0}}}.
\end{equation}
In the initial equilibrium state, any inertial effects in the gas phase can be neglected since $dR/dt|_{t=0}=0$.

For the present DPD simulations, the gas-liquid density ratio is not negligible, with $\rho_G / \rho_L \sim 0.2$.
Therefore, the motion in the gas phase must be taken into consideration.
In principle this is a compressible flow, but the time scale for pressure waves to traverse the bubble $R/c_G$, where $c_G$ is the sound speed in the gas phase, is very short compared to the other processes.
In other words, the Mach number $\epsilon = u(R,t)/c_G \ll 1 $. This aligns with other compressible RP models~\cite{prosperetti1986bubble,fuster2011liquid,wang2014multi}, where the primary focus is on pressure waves in the liquid at large distances from the bubble.
Under these conditions, the near-field remains essentially incompressible.
As the gas bubble expands, the initial approximation is that the local rate of expansion is uniform at all points within the bubble.
A radial potential flow, $\mathbf{u} = \nabla \phi$, is present, with the governing equations:
\begin{eqnarray}
   \nabla^2 \phi &=& \frac{2}{R} \frac{\mathrm{d} R}{\mathrm{d}t}, \\
   u(r,t) &=& \frac{r}{R} \frac{\mathrm{d} R}{\mathrm{d}t}.
\end{eqnarray}

Consequently, the density of the gas $\rho_G (t)$ is uniform within the bubble and can be expressed relative to the initial conditions:
\begin{equation}
  \rho_G(t) = \rho_{G0} (R_0 / R(t) )^2.
\end{equation}

The predominant pressure is mainly dictated by the thermodynamic pressure,denoted as $p_{B}(t)$, complemented by a minor correction arising from the fluid motion in the gas phase, expressed as $p_{1} (r, t)$.
The total gas pressure is given by $p_G(r,t) = p_B(t) + p_{1} (r, t)$.
These pressure variations, contingent upon $\epsilon \ll 1$, lack the magnitude to appreciably alter the gas density, and any density corrections would be of higher order in $\epsilon$.
The pressure variation is governed by the Navier-Stokes equation, in conjunction with a Stokes model for bulk viscosity, and is formulated as:
\begin{equation}
    p_1(r,t) =  f(t) - \frac{\rho_G(t)}{2R} r^2 \frac{\mathrm{d}^2R}{\mathrm{d}t^2},
\end{equation}
where $f(t)$ is an arbitrary function from the integration.
One could choose to set $p_1(0,t)=0$ and use the center of the bubble as the reference point.
Instead, we set the average value of $p_1$ within the bubble to be zero, leaving $p_B$ as the average pressure in the bubble.
This aligns with the approach taken in evaluating the gas pressure within the bubble in the DPD simulations.
The result is,
\begin{equation}
 0=\int_{0}^{R}2\pi rp_2(r,t)\mathrm{d}r = -\rho_1(t)\left ( \frac{1}{2R} \frac{\mathrm{d}^2R}{\mathrm{d}t^2 }  \right ) \frac{\pi R^4}{2}  + \pi R^2 f(t).
\end{equation}
Combining these outcomes yields a corrected estimate for the normal stress in the gas phase at the bubble surface:
\begin{equation}
\tau^B_{rr}(t) = -p_B(t) +\frac{1}{4} \rho_G (t) R  \frac{\mathrm{d}^2R}{\mathrm{d}t^2 } + 2 \mu_G \frac{1}{R}\frac{dR}{dt}.
\end{equation}

In conclusion, the 2D Rayleigh-Plesset equation, accounting for gas flow in the bubble, is given by:
\begin{eqnarray} \label{eq.full.2DRP}
    &&p_B(t) - \frac{1}{4}\rho_{G}(t) R \frac{\mathrm{d}^2R}{\mathrm{d}t^2} - p_{L}(R_E,t) \nonumber\\
    &&= \rho_L \log_e \left ( \frac{R_E}{R}  \right ) \frac{\mathrm{d}}{\mathrm{d}t}\left ( R \frac{\mathrm{d}R}{\mathrm{d}t}  \right ) -
    \frac{1}{2}\rho_L \left ( 1-\frac{R^2}{R_E^2} \right ) \left ( \frac{\mathrm{d}R}{\mathrm{d}t}  \right )^2
    + \left ( 2\mu_L + \frac{2}{3}\mu_G  \right ) \frac{1}{R} \frac{\mathrm{d}R}{\mathrm{d}t} + \frac{\gamma}{R}.
\end{eqnarray}
This equation is employed in our present study.
It is important to note that DPD simulations yield the average force on the upper and lower walls, acting as no-slip boundaries~\cite{2021_Lin_Operator_JCP}. Since these surfaces are flat, and the liquid is essentially incompressible, there is no normal viscous stress, and the average force corresponds to the liquid pressure.
Therefore, we omit consideration of the viscous normal stress at $r = R_E$ and directly relate $R(t)$ to $p_{E}(t)$.

\section{A Composite Neural Operator Network}\label{sec:3}
\begin{figure}[t]
    \centering
    \includegraphics[width=0.80\textwidth]{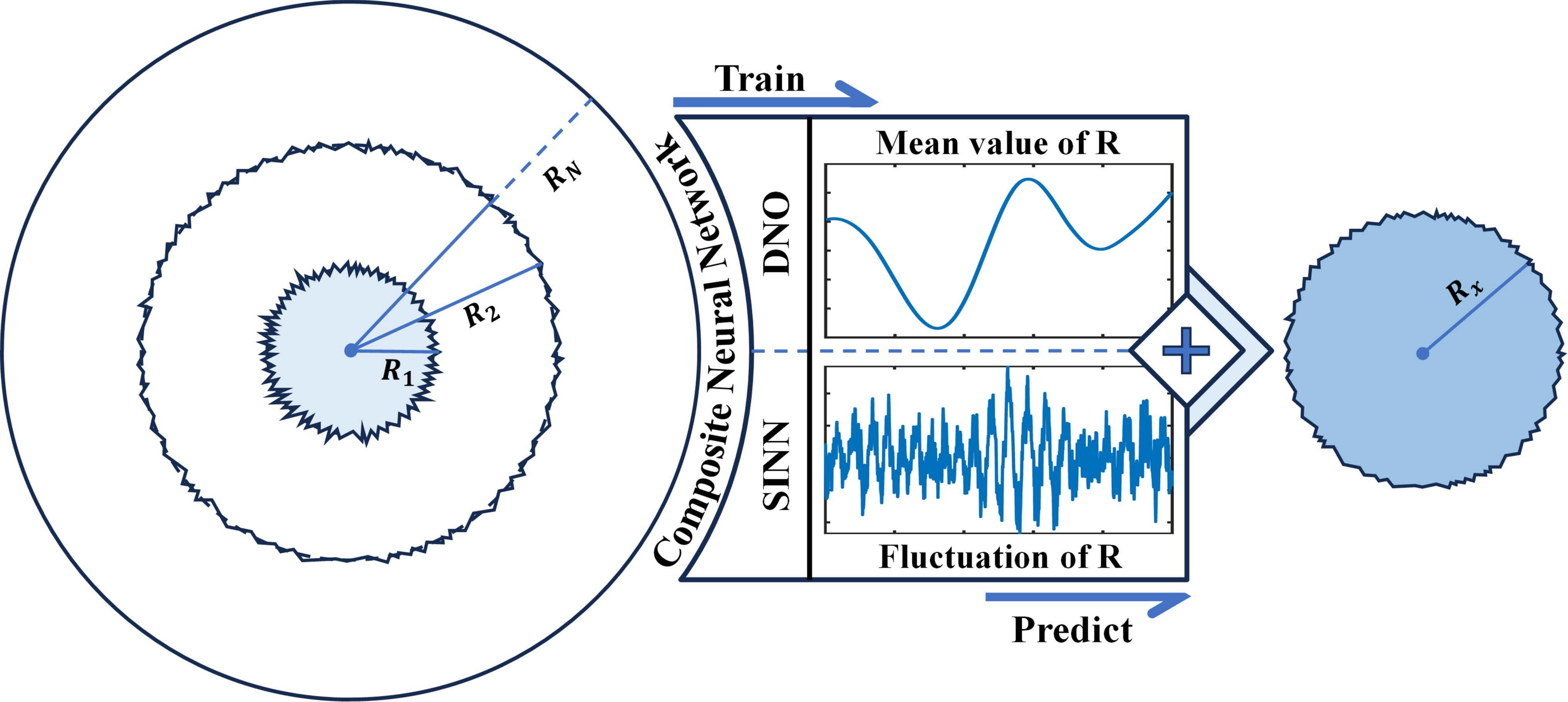}
    \caption{Schematic of the Composite Neural Operator Network. It contains a deep neural operators (DNO) network and a statistics-informed neural network (SINN). The entire network is trained on computational data generated from the continuum-based RP model and microscale DPD-mDPD simulations. A well-trained composite neural operator model is expected to provide correct predictions of bubble dynamics for both microscale and continuum scale.}
    \label{fig-composite_DNO}
\end{figure}

A composite neural operator network is proposed to learn multiscale bubble dynamics under pressure variations, which is designed to not only predict the mean behavior of the bubble dynamics from microscale to continuum regimes but also capture the correct stochastic fluctuations in microscale bubble dynamics. Figure~\ref{fig-composite_DNO} illustrates the overall diagram of the composite neural network. Due to the powerful learning ability for dynamic systems of Deep Neural Operators (DNO), it is used to approximate the nonlinear operator for the time evolution of the bubble radius $R(t)$ under a time-dependent external pressure $P_\infty(t)$ for both DPD and PR models. A statistics-informed neural network (SINN) constructed based on the long short-term memory (LSTM) architecture is used to learn the statistical features of the stochastic fluctuations in DPD bubble dynamics.

\subsection{Operator Learning for Mean Bubble Dynamics}\label{sec:3.1}
DNNs share a structure similar to artificial neural networks but usually with many hidden layers, providing them the ability to solve intricate implicit problems. For instance, a basic feed-forward neural network (FNN) has its final output through a combination of nonlinear and linear transformations applied to its original neural inputs. An $L$-layer FNN can be expressed as,
\begin{equation}
F(x)=G^{(L)}(\cdots (\sigma G^{(3)}(\sigma G^{(2)}(\sigma G^{(1)}(x))))),\label{eq1}
\end{equation}
where $G^{(*)}(x) = W^{(*)}x+b^{(*)}$ and $\sigma$ is the activation function. There are three nonlinear functions, i.e., Rectified Linear Activation (ReLU), Logistic (Sigmoid), and Hyperbolic Tangent (Tanh), are widely used as the activation function in deep learning~\cite{2022_Dubey_Activation_N}.
Regardless of a simple architecture, FNN has shown a great efficiency in solving many implicit problems. In this paper, FNN with a Tanh activation function is used to construct the deep learning framework for the DNO model.

\begin{figure}[t]
    \centering
   \includegraphics[width=0.48\textwidth]{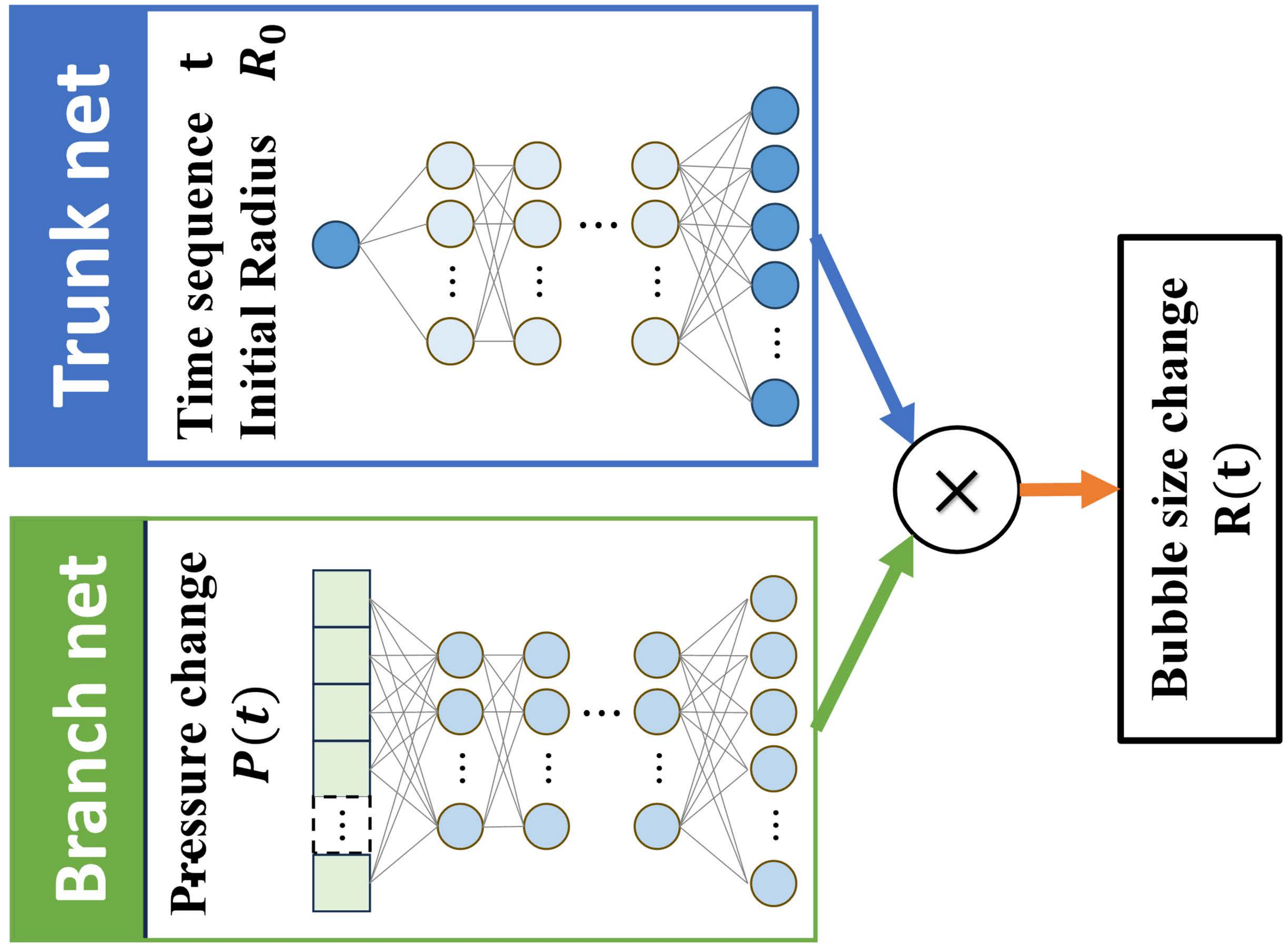}
    \caption{Schematic of the DNO model. The input of data, time series $t$ and time-dependent pressure $P(t)$, go through two neural networks, i.e., trunk net and branch net. The matrix product of the two neural network outputs becomes the final output of this DNO model.}
    \label{fig4}
\end{figure}

A schematic of the DNO model is presented in Fig.~\ref{fig4}, which is developed based on DeepONet by introducing a concept of sequence-to-sequence training. The DNO model consists of a DNN as the trunk net and another DNN as the branch net. The output of the DNO model is determined by the multiplication of the outputs from the trunk of branch nets, which is developed based on the universal approximation theorem of operators~\cite{2021_Lu_Learning_NMI}.
To speed up the training process, the trunk net takes the unchanged time sequence as its input, using the vector $\mathbf{t} = (t_{1}, t_{2}, t_{3}, ..., t_{n})$ with $n$ being the total number of discrete time points. The inputs of branch net are the corresponding sequences $Y$ related to this time sequence $t$, using the vector $Y_{i} = (Y_{i}(t_{1}), Y_{i}(t_{2}), Y_{i}(t_{3}), ..., Y_{i}(t_{n}))$ with $i\in[0, m]$ being the $i$-th training dataset and $m$ being the total number of training datasets.
The entire sequence $Y_{i}$ is fed to the neural networks to train the DNO model. Therefore, instead of one-to-one training, the training process is finished by sequence-to-sequence and thus shorten the training time effectively, which has been verified by Lu et al.~\cite{2023_Lu_Deep_CM}. Moreover, the setting of the trunk net ensures the continuity of the data  and makes it easier to display the time and space-dependent features, which leads to higher accuracy in operator learning.

Both the trunk net and the branch net of the DNO framework are $L$-layer FNN with a Tanh activation function. The trunk net has one neuron in the input layer to take time vector and the initial bubble size $R_0$ and $n$ neurons in the output layer. The output of trunk net is a $n\times n$ matrix which can be expressed as $F_{T}(\mathbf{t})=F(\mathbf{t})$.
For the branch net, it has $m$ neurons in the input layer, $n$ neurons in the output layer. The output of branch net is a $m\times n$ matrix and can be expressed as $F_{B}(\mathbf{t})=F(Y_{i}(\mathbf{t}))$. Therefore, as illustrated in Figure~\ref{fig4}, the final output of the DNO model is $R_{i}(\mathbf{t}) = F_{B}(\mathbf{t})\cdot F_{T}(\mathbf{t})$ with a dimension of $m\times n$.

\subsection{SINN for Correlated Fluctuations}\label{sec:3.2}
\begin{figure}[t]
    \centering
    \includegraphics[width=0.75\textwidth]{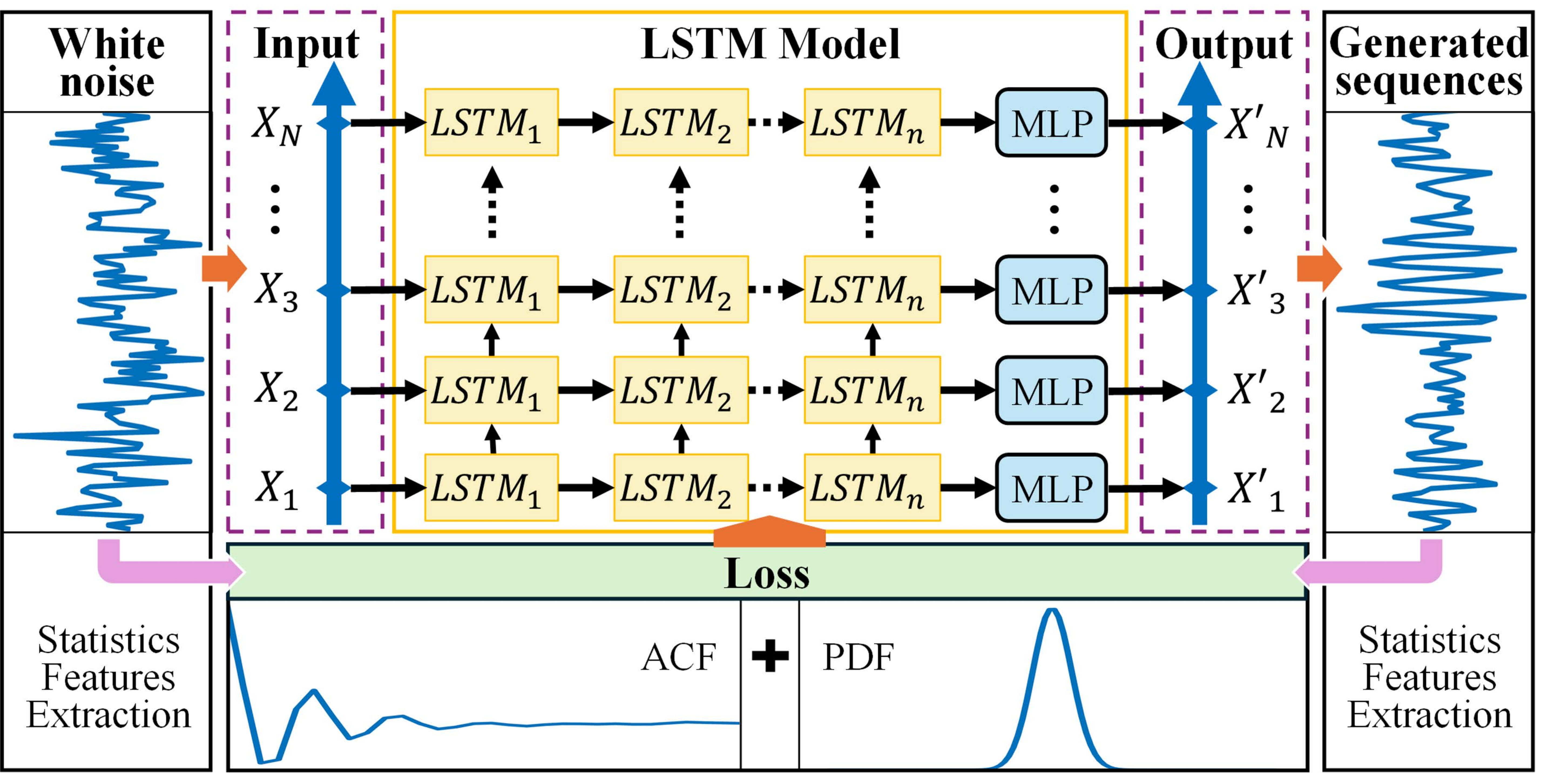}
    \caption{Schematic of SINN. It demonstrates the process in which white noise evolves into a target trajectory through multiple layers of LSTM cells. The SINN doesn't directly learn the exact values of the target but continuously learns the statistical features of the target to achieve the final trajectory generation.}
    \label{fig-SINN_structure}
\end{figure}

The statistics-informed neural network (SINN) is used to learn the non-Markovian stochastic dynamics from data~\cite{2023_Zhu_Learning}. It is proved that it can be used as a universal approximator for stochastic dynamics. The structure of the SINN model is shown in Figure~\ref{fig-SINN_structure}, where the long short-term memory (LSTM) architecture is used to capture non-Markovian memory effects of a stochastic process with correlated fluctuations.

The input white noise goes through the multi-layer LSTM, a dense layer and turns into stochastic time series. The process of white noise sequence $\xi_t$ passing the $n$-th layer of LSTM can be expressed as,
\begin{align}
f_t^{(k)} &= \sigma_g (W_f \xi_t + U_f h_{t-1}^{(k)} + b_f),\nonumber \\
i_t^{(k)} &= \sigma_g (W_i \xi_t + U_i h_{t-1}^{(k)} + b_i),\nonumber \\
o_t^{(k)} &= \sigma_g (W_o \xi_t + U_o h_{t-1}^{(k)} + b_o),\\
\tilde{c}_t^{(k)} &= \sigma_c (W_c \xi_t + U_c h_{t-1}^{(k)} + b_c),\nonumber \\
c_t^{(k)} &= f_t^{(k)} \circ c_{t-1}^{(k)} + i_t^{(k)} \circ \tilde{c}_t^{(k)},\nonumber \\
h_t^{(k)} &= o_t^{(k)} \circ \sigma_h (c_t^{(k)}),\nonumber
\end{align}
where $i_t^{(k)}$ is the input gate, $f_t^{(k)}$ is the forget gate, and $o_t^{(k)}$ is the output gate of the $k$-th layer, $k=1,2,3,...,\rm{n}$. $c_t^{(k)}$ is the cell state and $\tilde{c}_t^{(k)}$ is the cell input activation. $\sigma_*$ is activation function and $W_*$, $b_*$ are hyperparameters. The final output $\chi_t$ can be written as $\chi_t = W_\chi h_t^{(n)}$.

To generate the stochastic time series which have the same statistical features of the target time series, the probability density function (PDF) and the autocorrelation function (ACF) are selected as two important statistical information for the SINN model to learn. PDF is a statistical function that describes the likelihood of a continuous random variable falling within a particular range of values. It ensures that the generated time series has a similar amplitude to the target sequence.
In numerical computation, due to the discrete nature of histogram operations, binning-based probability density function estimators are not differentiable. Then, the kernel density estimation (KDE) can be used to compare the empirical PDFs of both the target and generated trajectories. The kernel density $f$ at any given point $x$ can be expressed as,
\begin{equation}
f_h(x) =  \frac{1}{N} \sum_{{i=0}}^{{N}} K_h (x-x_i),
\label{KDE}
\end{equation}
where $K_h(x)=K(x/h)/h$ with $K$ being a non-negative kernel and $h$ being a positive bandwidth. Here, the Gaussian kernel $K^{\rm Gauss}(x)=\exp(-x^2/2)/\sqrt{2\pi}$ with $h=N^{-1/5}$ is used.

In addition to ensuring the congruence of the PDFs between the target data and the generated data, it is essential to incorporate the extraction of stochastic process memory information to enhance the fitting of statistical characteristics to the target. ACF provides insights into how the correlation between two values of a sequence evolves with changes in their temporal separation~\cite{2000_Mohamed_Data}. It quantifies the self-similarity within a signal at various time delays. The ACF of a discrete-time sequence $\mathbf{x}$ at lag $\tau$ is defined as,
\begin{equation}
{\rm ACF}_x(\tau) = \lim_{{N \to \infty}} \frac{1}{(N-\tau)\cdot\sigma_x^2} \sum_{{n=1}}^{{N-\tau}} \left[x_n-\Bar{x}\right] \cdot \left[x_{n + \tau}-\Bar{x}\right],
\label{ACF}
\end{equation}
where $N$ is the size of dataset $\mathbf{x}$, $\sigma_x$ is the standard deviation of $\mathbf{x}$ and $\Bar{x}$ is the mean.
To enable the model to learn the statistical characteristics of the target sequence, the $L_2$ norm is used to define the loss function in the form of
\begin{equation}
{\rm Loss} = \| {\rm KDE}^G - {\rm KDE}^T \|_2 + \| {\rm ACF}^G - {\rm ACF}^T \|_2,
\label{loss}
\end{equation}
where ${\rm KDE}^G$, ${\rm ACF}^G$ are the KDE and ACF of the output generated sequences, and ${\rm KDE}^T$, ${\rm ACF}^T$ are the KDE and ACF of the target sequences, respectively.

\begin{figure}[ht]
    \centering
    \includegraphics[width=0.85\textwidth]{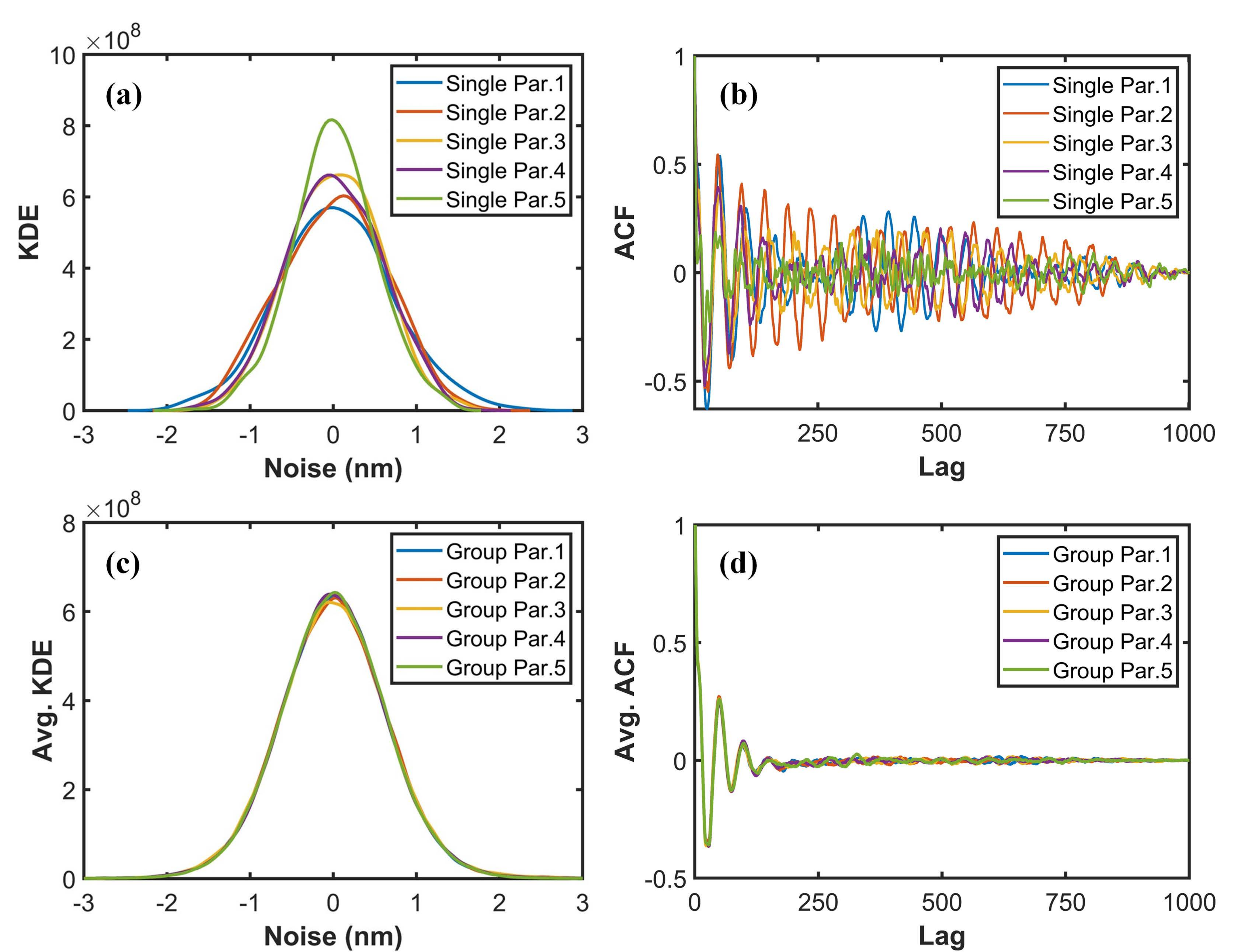}
    \caption{Statistical analysis of simulation results. (a) KDE of case \#1 results from different parallel runs. (b) ACF of case \#1 results from different parallel runs. (c) Averaged KDE of all cases from different parallel runs. (d) Ensemble-averaged ACFs of all cases from different parallel runs.}
    \label{fig-statistic}
\end{figure}

\section{Results}\label{sec:4}
To study full-scale bubble growth dynamics from microscale to continuum scale, it is necessary to generate bubble dynamics data at different length scales for training and testing the composite neural operator model. Dynamic data at the macroscopic scale is produced using the two-dimensional RP model for a gas bubble within a confined liquid medium with a finite gas-liquid density ratio. Microscopic bubble dynamics data is generated by the DPD-mDPD coupled simulation system, where the gas phase is simulated by DPD particles and the liquid phase is simulated by mDPD particles. The details of the simulations can be found in our previous work by Lin et al.~\cite{2021_Lin_Operator_JCP} with DPD parameters listed in its Table II. The liquid pressure is subjected to a Gaussian random process with an exponential function mask where the mean of variation $\mu=0~{\rm Pa}$ and the standard deviation $\sigma=9.76 \times 10^5~{\rm Pa}$. One thousand sets of pressure variation data are generated and fed into the the composite neural operator model. Among these, 500 sets serve as input for the 2D RP model at different initial radius states $R_0$ from $1~\rm{\mu m}$ to $10~\rm{\mu m}$, while the remaining 500 data sets are designated for the DPD-mDPD model at different initial bubble radius $R_0$ ranging from $100~\rm{nm}$ to $1~\rm{\mu m}$. Finally, 1000 sets of outputs corresponding to the time evolution of the bubble radius under pressure variations are obtained from the RP and DPD-mDPD simulations.
%

At the macroscopic scale, the 2D RP model characterizes the variation in radius $R(t)$ of the single circular gas bubble in response to changes in liquid pressure as shown in Figure~\ref{fig-simulation-setup}(d). In the simulation results at the microscopic scale, it can be observed that in addition to bubbles oscillating with dynamic pressure changes, there are also fluctuations present in the behavior of bubble size changes as shown in Figure~\ref{fig-simulation-setup}(c). The fluctuation can be attributed to the inherent stochastic nature of dynamic processes. This inherent randomness is a characteristic feature of systems at the microscopic scale, reflecting the complex and unpredictable nature of individual particle dynamics and interactions.
To analyze the characteristics of bubble fluctuations, 12 parallel DPD-mDPD simulations under identical initial conditions are conducted. From this, the mean variation of bubble size over time for each scenario can be calculated and subsequently, the noise variation for $i$-th case in the $n$-th parallel simulation, $R_{N}(t)_{i,n}$ can be obtained which can be written as,
\begin{equation}
R_{N}(t)_{i,n} = R(t)_{i,n}-\frac{1}{12} \sum_{{n=1}}^{{12}} R(t)_{i,n},
\label{Noise}
\end{equation}
where, $R(t)_{i,n}$ represents the bubble changes of $i$-th case in the $n$-th parallel simulation. In the results obtained through parallel simulations, with identical initial $R0$ and the same pressure changes, the noise terms are different. Similarly, the KDE and ACF of the noise under different parallel runs also exhibit variations. Figure~\ref{fig-statistic} (a), (b) respectively display the KDE for $R_{N}(t)_{1,1}, R_{N}(t)_{1,2}, R_{N}(t)_{1,3}, R_{N}(t)_{1,4}, R_{N}(t)_{1,5}$, as well as ACF for the same noise data.
This observation indicates that if the model is only learning a specific noise term or the ACF and KDE information from a single parallel run, it lacks generalization across potential noise terms. However, the KDEs of the combined results for all different scenarios within each parallel run are nearly consistent. The ensemble-averaged ACFs across each parallel run are close. Therefore, in the training process of the composite neural operator network, the DNO model is required to learn the dynamic variations of the mean value of the bubble radius. Simultaneously, the SINN model captures the overall KDE and averaged ACF. The trained SINN, serving as a generative model, can effectively produce random sequences that align with the original statistical features of the noise. This comprehensive approach enables the prediction of the entire multiscale bubble evolution.

\begin{figure}[t]
    \centering
    \includegraphics[width=0.85\textwidth]{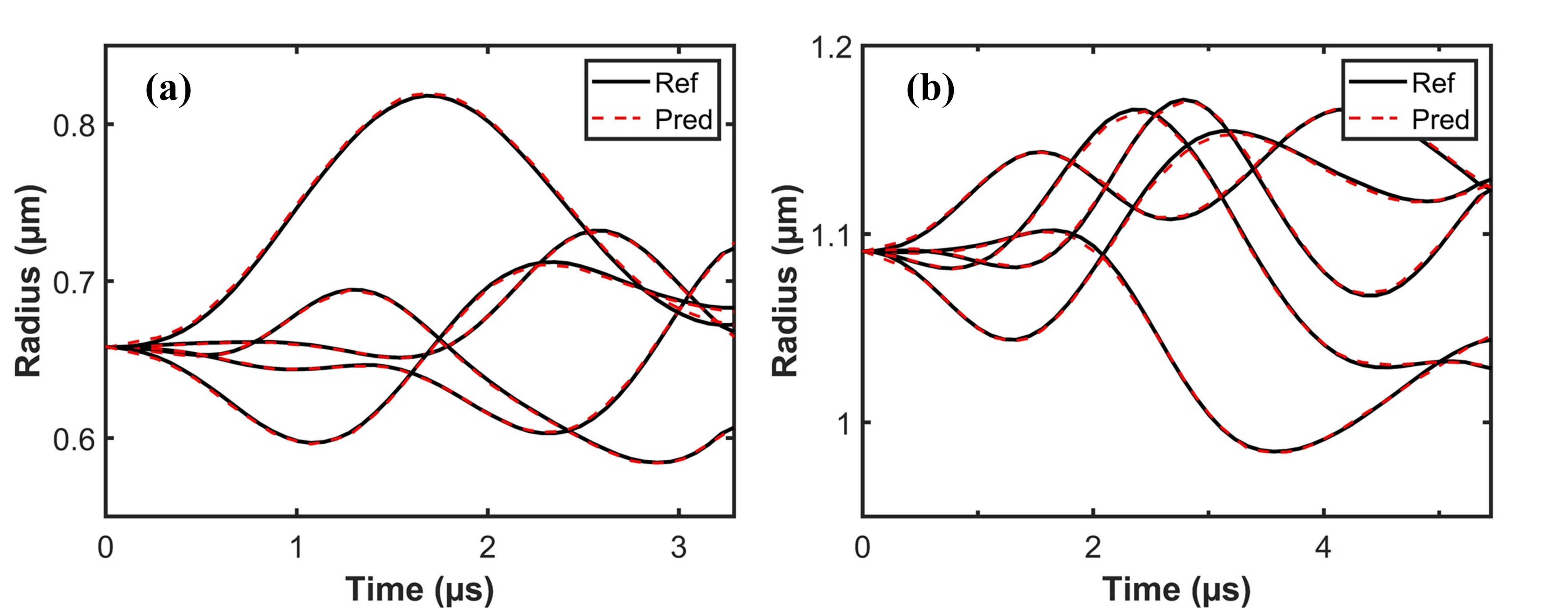}
    \caption{Mean value prediction results. (a) shows 5 test cases of bubble dynamic prediction at initial $R_0 = 658~\rm{nm}$ with reference data from DPD-mDPD simulation. (b) shows 5 test cases of bubble dynamic prediction at initial $R_0 = 1.091~\rm{\mu m}$ with reference data from 2D RP model.}
    \label{fig-pred_mean}
\end{figure}

The composite neural operator network is trained based on the generated datasets from the simulation. 400 datasets from the 2D RP model, 400 datasets from the DPD-mDPD simulation are used for training, and the rest are used for testing. Figure~\ref{fig-pred_mean} displays 10 prediction results to (a) showcases the predicted variations in bubble radius under five different pressure changes when the initial $R_0 = 658~\rm{nm}$, and (b) illustrates the predicted variations in bubble radius under five different pressure scenarios when the initial $R_0$ is set to $1.091~\rm{\mu m}$. The prediction accuracy (Acc) is described by,
\begin{equation}
{\rm Acc} = 1 - \left\|\frac{{\rm Pred}-{\rm Ref}}{\rm Ref}\right\|_2,
\label{Noise}
\end{equation}
where $\|*\|_2$ represents the $L_2$-norm between the predicted (Pred) and reference (Ref) values. In both cases, the dynamic changes in the bubble radius do not exhibit significant noise fluctuations. The model can accurately predict the dynamic evolution process with an accuracy of over 99\%.

\begin{figure}[t]
    \centering
    \includegraphics[width=0.85\textwidth]{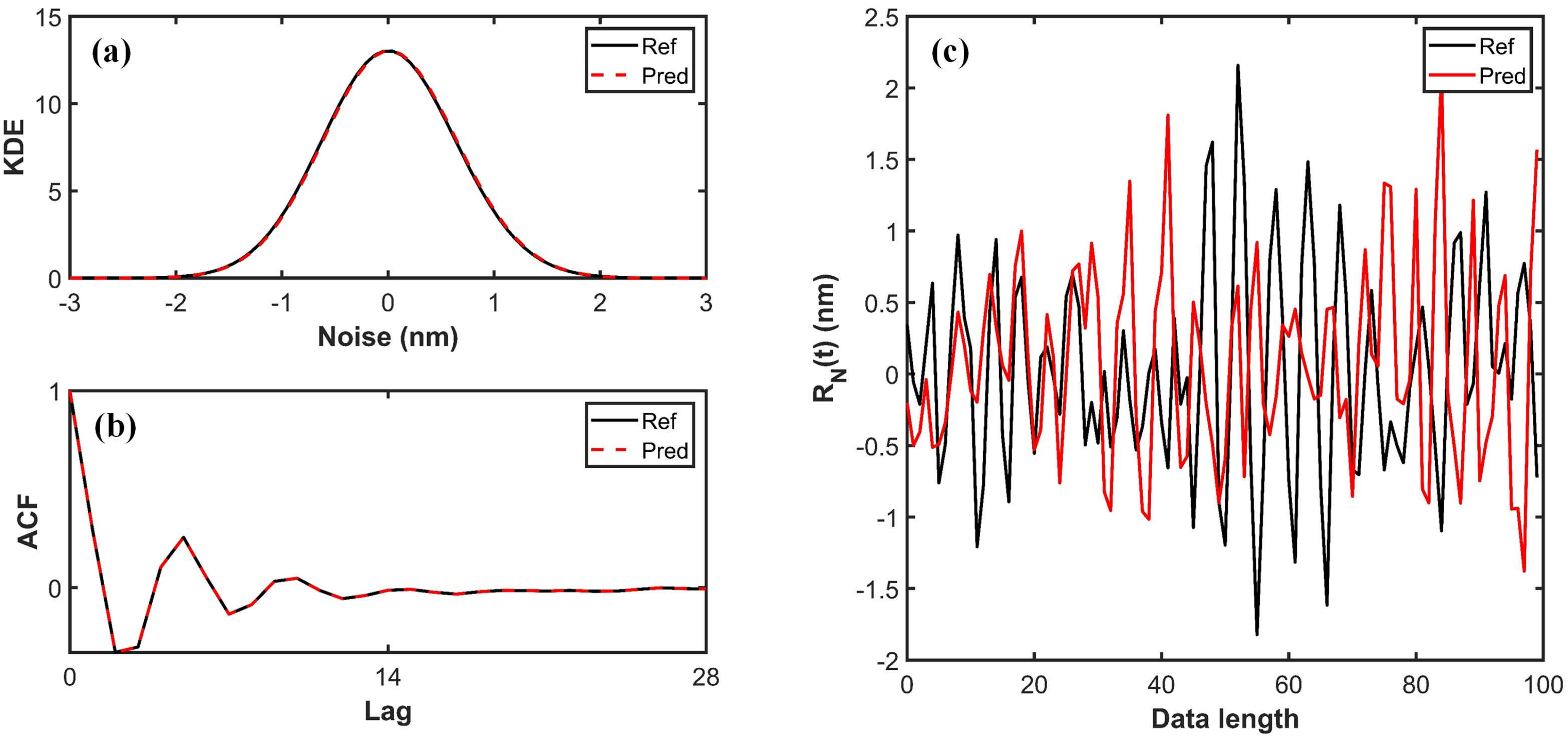}
    \caption{Generated model results. (a) shows the KDE training results. (b) shows the ACF training results. (c) shows one generated noise trajectory from the trained model compared with the reference noise trajectory.}
    \label{fig-SINN_prediction}
\end{figure}

As $R_0$ gradually decreases, the noise in the bubble evolution process gradually becomes apparent. The noisy data is derived from simulation data of DPD-mDPD under the condition where $R_0$ is near the lower boundary. The overall KDE and ensemble-averaged ACF can be found and set as training targets. The SINN part of the composite neural operator network acts as a generative model to create new noise trajectories that satisfy the same KDE and ACF of the ensemble noise data. Figure~\ref{fig-SINN_prediction} shows the result of one case of the generated trajectory. It can be observed that the model well learns the statistical characteristics (KDE and ACF) of the noise, and generates a brand-new noise trajectory.

\begin{figure}[th]
    \centering
    \includegraphics[width=0.95\textwidth]{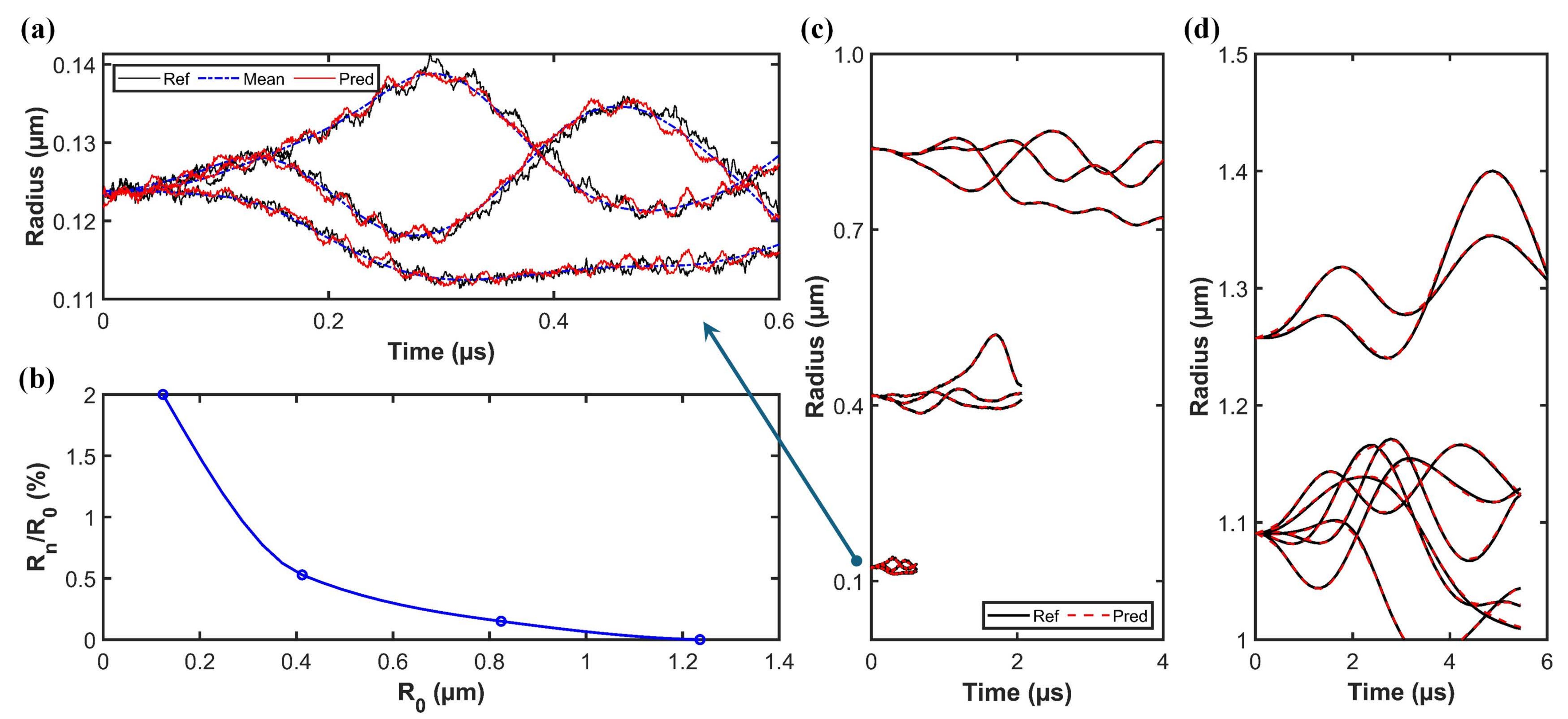}
    \caption{Bubble dynamic prediction. (a) One example case of model prediction results at the microscale. (b) The relationship between $R_N/R_0$ and $R_0$. (c) Model prediction of bubble radius within $1~\rm{\mu m}$ (d) Model prediction of bubble radius over $1~\rm{\mu m}$.}
    \label{fig-predict_all_total}
\end{figure}
Subsequently, by integrating the accurate predictive ability of DNO for the mean value, the entire model can capture the dynamic fluctuations in bubble behavior at the microscale as shown in Figure~\ref{fig-predict_all_total} (a). The predicted results of the model, incorporating statistical features, encompass not only a singular mean prediction but also include a fluctuation term, capturing the dynamic variations in bubble behavior at the microscale. It is noteworthy that the fluctuation terms in the dynamic changes of bubble growth under identical conditions are different. Therefore, the predicted data may not align perfectly with the reference data.

Through the analysis of simulation data, the relationship between $R_n$ and $R_0$ can be determined. Figure~\ref{fig-predict_all_total} (b) illustrates the variation of $R_N/R_0$ with respect to $R_0$. From a data processing perspective, as $R_0$ continuously increases, the simulation scale gradually expands. The statistical features manifested by the fluctuation term of bubble growth become challenging to express in the data. The absolute value of the $R_N$ tends to decrease as $R_0$ increases. Consequently, the $R_N/R_0$ curve initially experiences a rapid decline. After $R_0$ exceeds $0.7~\rm{\mu m}$, the ratio of the fluctuation term $R_N$ to $R_0$ is less than 0.22\%. This explains that beyond this point, the RP model and DPD model will converge, yielding identical results. Incorporating this relationship into the composite neural operator network allows it to adapt to any initial $R_0$, predicting the corresponding dynamic bubbles growth under any pressure changes. Figure~\ref{fig-predict_all_total} (c) shows the prediction results with bubble radius between $0.1~\rm{\mu m}$ to $1.0~\rm{\mu m}$. And Figure~\ref{fig-predict_all_total} (d) shows the prediction results with bubble radius between $1.0~\rm{\mu m}$ to $1.5~\rm{\mu m}$.

The results demonstrate that the model effectively predicts bubble dynamics across different scales and captures the variations in the fluctuation term at a microscopic level. Therefore, the well-trained composite neural operator network seamlessly integrates the multiscale aspects, eliminating the need to switch models when addressing problems at different scales. It can serve as a powerful and unified surrogate model for tackling diverse scale-related problems.

\section{Summary and Discussion}\label{sec:5}
\vspace{-0.13cm}
A composite neural operator model for multiscale bubble dynamics has been developed to unify microscale stochastic bubble dynamics models and macroscale continuum-based bubble dynamics models, which is able to accurately approximate both the deterministic aspects of bubble dynamics and their stochastic fluctuations at microscale due to fluctuating hydrodynamics. This is realized through the integration of a deep neural operator for predicting the mean bubble dynamics and a statistics-informed neural network for capturing correlated fluctuations by leveraging a long short-term memory architecture to consider the non-Markovian effects.

We performed many-body dissipative particle dynamics simulations for stochastic bubble growth dynamics under pressure variations to generate training and testing data at the microscale regime, and solved the continuum-based Rayleigh-Plesset equation for deterministic bubble growth dynamics under pressure variations to generate training and testing data at macroscale regime. The effectiveness of the composite neural operator model has been demonstrated through comparison of predictions with numerical simulations from microscopic to macroscopic scales, showing a high accuracy in predicting bubble dynamics under varying conditions. The model successfully reconciles the deterministic and stochastic realms of bubble behavior, offering a holistic view of bubble dynamics that was previously unattainable with traditional standalone physics-based bubble dynamics models.

The composite neural operator developed in this work is the first deep learning surrogate for multiscale bubble growth dynamics that can capture correct stochastic fluctuations in microscopic fluid phenomena. At the microscale, the nonlinear dynamics of micro-bubble can be significantly affected by the thermo-hydrodynamic fluctuations at the liquid-vapor interface, which refers to the energy changes and structural disturbances at the bubble surface induced by the microscale fluctuations of molecular motion. The stochastic fluctuation plays a crucial role in the mass exchange, energy transfer, and interfacial chemical reaction between the micro-bubble and the surrounding fluids.
There are a lot of important chemicophysical processes in industrial applications, i.e., surface adsorption, catalytic reaction, mass transfer, heavily rely on the interactions between micro-bubbles and surrounding fluids~\cite{2022_Sakr_ACritical_AEJ}. Examples include the heat/mass transfer enhancement technology using micro-bubbles formed on the surface of porous media to promote the efficiency of catalytic reactions~\cite{2023_Jung_Industrial_ACIS}, and the proton exchange membrane fuel cells using micro-bubbles for hydrogen gas dispersion to accelerate the hydrogen oxidation reaction~\cite{2024_Stoll_Impacts_JPS}.

This research emphasized the importance of integrating multiscale fluid physics to tackle the complex nature of nonlinear bubble dynamics across scales. By bridging the gap between microscale stochastic fluid models and macroscale continuum-based fluid models, it provides a comprehensive understanding of bubble behavior at different scales, which is crucial for optimizing relevant industrial processes involving multiscale bubble dynamics. Furthermore, this study highlights the potential of advanced computational methods, such as deep neural networks and neural operator learning, in enhancing our understanding of complex fluid phenomena, setting a new benchmark for future investigations in the field of multiscale fluid dynamics.
\vspace{-0.05cm}

\section*{Acknowledgments}
\vspace{-0.15cm}
This work was supported as part of the AIM for Composites, an Energy Frontier
Research Center funded by the U.S. Department of Energy, Office of Science, Basic Energy
Sciences at Clemson University under award \#DE-SC0023389.

\end{document}